\documentclass[namedreferences]{solarphysics}

\usepackage[hyperref,optionalrh,showbiblabels]{spr-sola-addons} 
\usepackage{epsfig}          
\usepackage{epstopdf}
\usepackage{graphicx}        
\usepackage{color}           
\usepackage{breakurl}        

\chardef\us=`\_

\begin{document}

\begin{article}
\begin{opening}

\title{Observations of a Radio-quiet Solar Preflare\\
}

\author[addressref={aff1,aff2},corref,email={benz@astro.phys.ethz.ch}]{\inits{A.O.}\fnm{Arnold O.}~\lnm{Benz}\orcid{0000-0001-9777-9177}}
\author[addressref=aff1,email={marina.battaglia@fhnw.ch}]{\inits{M.}\fnm{Marina}~\lnm{Battaglia}\orcid{0000-0003-1438-9099}}
\author[addressref=aff3,email={manuel.guedel@univie.ac.at}]{\inits{M.}\fnm{Manuel}~\lnm{G\"udel}}


\address[id=aff1]{Institute for 4D Technologies, FHNW, 5210 Windisch, Switzerland}
\address[id=aff2]{Institute for Astronomy, ETH Zurich, 8093 Z\"urich, Switzerland}
\address[id=aff3]{University of Vienna, Department of Astrophysics, T\"urkenschanzstrasse 17, A-1180 Vienna, Austria}

\runningauthor{A.O. Benz {\it et al.}}
\runningtitle{Electron Acceleration in a Solar Preflare}

\begin{abstract}
The preflare phase of the flare SOL2011-08-09T03:52 is unique in its long duration, its coverage by the {\it Reuven Ramaty High Energy Solar Spectroscopic Imager (RHESSI) and the Nobeyama Radioheliograph}, and the presence of three well-developed soft X-ray (SXR) peaks. No hard X-rays (HXR) are observed in the preflare phase. Here we report that also no associated radio emission at 17 GHz was found despite the higher sensitivity of the radio instrument. The ratio between the SXR peaks and the upper limit of the radio peaks is larger by more than one order of magnitude compared to regular flares. The result suggests that the ratio between acceleration and heating in the preflare phase was different than in regular flares. Acceleration to relativistic energies, if any, occurred with lower efficiency.
\end{abstract}
\keywords{Flares, Pre-Flare Phenomena; Energetic Particles, Acceleration; Radio Bursts, Microwave (mm, cm); X-Ray Bursts, Soft}
\end{opening}

\section{Introduction}
     \label{S-Introduction}

Solar flares are generally assumed to be the result of magnetic reconnection in the corona. MHD describes the build up and storage of free magnetic energy, which is then released impulsively \citep[{\it e.g.} reviews by][]{2011LRSP....8....6S,2014masu.book.....P}. However, the actual energy release is not understood. A substantial fraction of the energy is initially converted to accelerate energetic (non-thermal) electrons and ions. The energy content in flare-accelerated particles equals the total bolometric radiation within the error margin \citep[factor of two,][]{2012ApJ...759...71E} and exceeds the SXR emission in  most cases \citep{2017ApJ...836...17A}. The bolometric radiation is approximated by all emissions from X-rays to infrared. Most of the flare radiation is emitted by thermal plasmas at temperatures from $10^4$K (white light) to $10^7$K (SXR). The near-equality of particle and thermal energy is consistent with the hypothesis that most flare energy is first released into particle acceleration and subsequently thermalized. Several acceleration processes have been proposed, but there is no general consensus \citep{2011SSRv..159..357Z}. Thus flares must be considered primarily as unknown particle accelerators.

The energy radiated in SXR amounts to 20\% of the bolometric emission on the average in large flares \citep{2012ApJ...759...71E}. Again, the error margin of individual events is at least a factor of two, and their range is 9\,--\,63 \%.

Most of the energy of non-thermal electrons is in the rage of 10\,--\,30 keV particles. They emit HXR bremsstrahlung at similar photon energies. Signatures of non-thermal electrons are also observed in radio waves \citep[{\it e.g.} review by][]{1998ARA&A..36..131B} and HXR \citep[{\it e.g.}  review by][]{2011SSRv..159..421L}. Gyro-synchrotron emission (above about 3 GHz) originates from mildly relativistic electrons ($>$ 100 keV). HXR emission is dominated by bremsstrahlung of electrons in the range 10\,--\,30 keV. The radio emissions at 17 GHz and HXR $>$ 30 keV are well correlated \citep{1988ApJ...324.1118K}, indicating the high-energy and low-energy electrons are closely related. In some cases, large radio telescopes are more sensitive in detecting non-thermal electrons than current HXR detectors, although the radio-emitting particles have a much higher energy. Thus optically thin gyro-synchrotron emission has been used as a tracer for flare-accelerated electrons in stellar flares and weak solar events. However, the fraction of electrons accelerated to relativistic energies is an open question and is related to the acceleration process.

The ratio between gyro-synchrotron emission and thermal SXRs in flares was found to be stable over seven orders of magnitude of solar and stellar flare energies \citep{1993ApJ...405L..63G,1994A&A...285..621B}. The two emissions are related as
\begin{equation}
  \log {\left({L_{\rm X}\over L_{\rm R}}\right)} = 15.5 \pm 0.5,
\end{equation}
where $L_{\rm X}$ [erg s$^{-1}$] is the luminosity in the SXR band, $L_{\rm R}$ [erg s$^{-1}$ Hz$^{-1}$] is the radio luminosity density at 8\,--\,9 GHz, and $\pm$ indicates the width of the distribution. The correlation is equally good for time-integrated flare luminosities. The invariability suggests that small and large flares have similar ratios between thermal and non-thermal energies and capabilities to accelerate relativistic electrons, and also have comparable magnetic fields in the source regions of the gyro-synchrotron emission. The most prominent deviation was noted in microflares, which were found to be relatively weak in radio emission \citep{1994A&A...285..621B}. The same trend was found to be even more pronounced in quiet-region microflares \citep[also known as nanoflares,][]{2000SoPh..191..341K}, which show reduced radio luminosities compared to the above relation by an order of magnitude.

What is the ratio of gyro-synchrotron and SXR emission in the build-up phase of a large flare? In the preflare phase the coronal plasma slowly heats up until the HXR emission suddenly and impulsively increases \citep{2017LRSP...14....2B}. The preflare phase and impulsive flare phase are often well pronounced (see, {\it e.g.}, Figure \ref{light curve}, top). Weak HXR emission is sometimes observed in preflares \citep{1983SoPh...83..267B,2009ApJ...705L.143S}. \citet{2009A&A...498..891B} reported four preflares with only thermal X-ray emission in RHESSI observations. However, \citet{2012ApJ...758..138A} searched for non-thermal emission of these events in more sensitive radio observations and found non-thermal radiation in two of these preflares using the Nobeyama Radioheliograph at 17 GHz and other radio instruments. This indicates that sometimes electrons are accelerated to relativistic velocities already in the build-up phase of flares.

Here we present observations of the early phase of a flare with 3 preflare peaks observed in SXR but with no associated HXR emission and search for corresponding peaks in radio emission. The goal is to study the efficiency of relativistic electron acceleration at the very beginning of solar flares.

\section{Observations} 
      \label{S-observations}

 The M2.5 flare SOL2011-08-09T03:52 was preceded by a prolonged preflare phase, lasting some 20 minutes from 03:00 to 03:21 UT. \citet{2014ApJ...789...47B} presented a multi-wavelength study of the event. They find that the non-thermal component of the 17 GHz radio flux rises simultaneously with the non-thermal 25\,--\,40 keV HXR emission only after the preflare phase and suggest ``that particle acceleration was minimal during the preflare phase." The background radio flux at 17 GHz can be interpreted as free-free emission from a thermal plasma of the active region with a temperature of 2 MK and a density of $\approx 1.2 \times 10^{10}$cm$^{-3}$ in agreement with SXR and EUV observations.

 Here we study the relation of SXR and gyro-synchrotron emission of the preflare in a different way, making use of the rare property that the selected long-duration preflare has three significant SXR enhancements. Such features are uncommon since the thermal emission is often found to correlate with the time-integrated flux of non-thermal electrons during the main phase \citep{1968ApJ...153L..59N, 1993SoPh..146..177D}. Thus SXR peaks only if the cooling of the hot plasma is faster than the heating. This is infrequent in the preflare phase. On the contrary, the emission measure and temperature of the hot plasma usually increase during the preflare phase, which may hide potential SXR peaks.

 \subsection{SXR Data and Analysis} 
  \label{SXR-text}

  \begin{figure}    
   \centerline{\includegraphics[width=0.9\textwidth,clip=1]{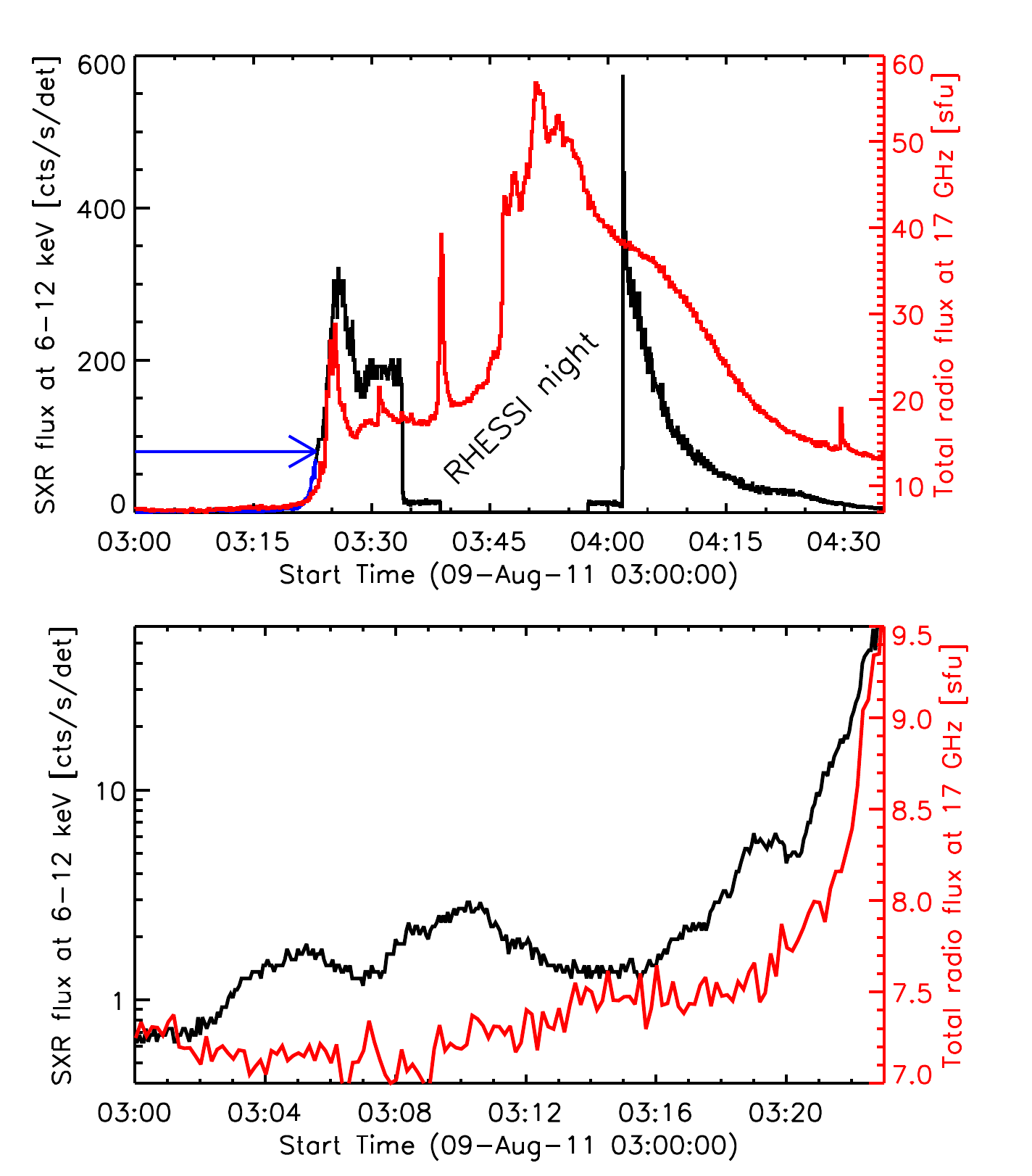} }
              \vskip0.5cm
              \caption{{\it Top:} Light curves of flare SOL2011-08-09T03:52 as observed in SXR (6\,--\,12 keV) by RHESSI ({\it black}) and the Nobeyama Radioheliograph ({\it red}). {\it Bottom:} Enlargement of the preflare phase which is indicated on the top with a blue arrow.}
   \label{light curve}
   \end{figure}

  We use SXR data from the {\it Ramaty High Energy Solar Spectroscopic Imager} \citep[RHESSI;][]{2002SoPh..210....3L}. The RHESSI spectra were fitted using OSPEX with a single thermal component and emission measure. No statistically significant non-thermal component was found in the preflare phase (before 03:21 UT) and the RHESSI images of the preflare emission show a single source at all times and energies.

\begin{figure}    
   \centerline{\includegraphics[width=0.9\textwidth,clip=1]{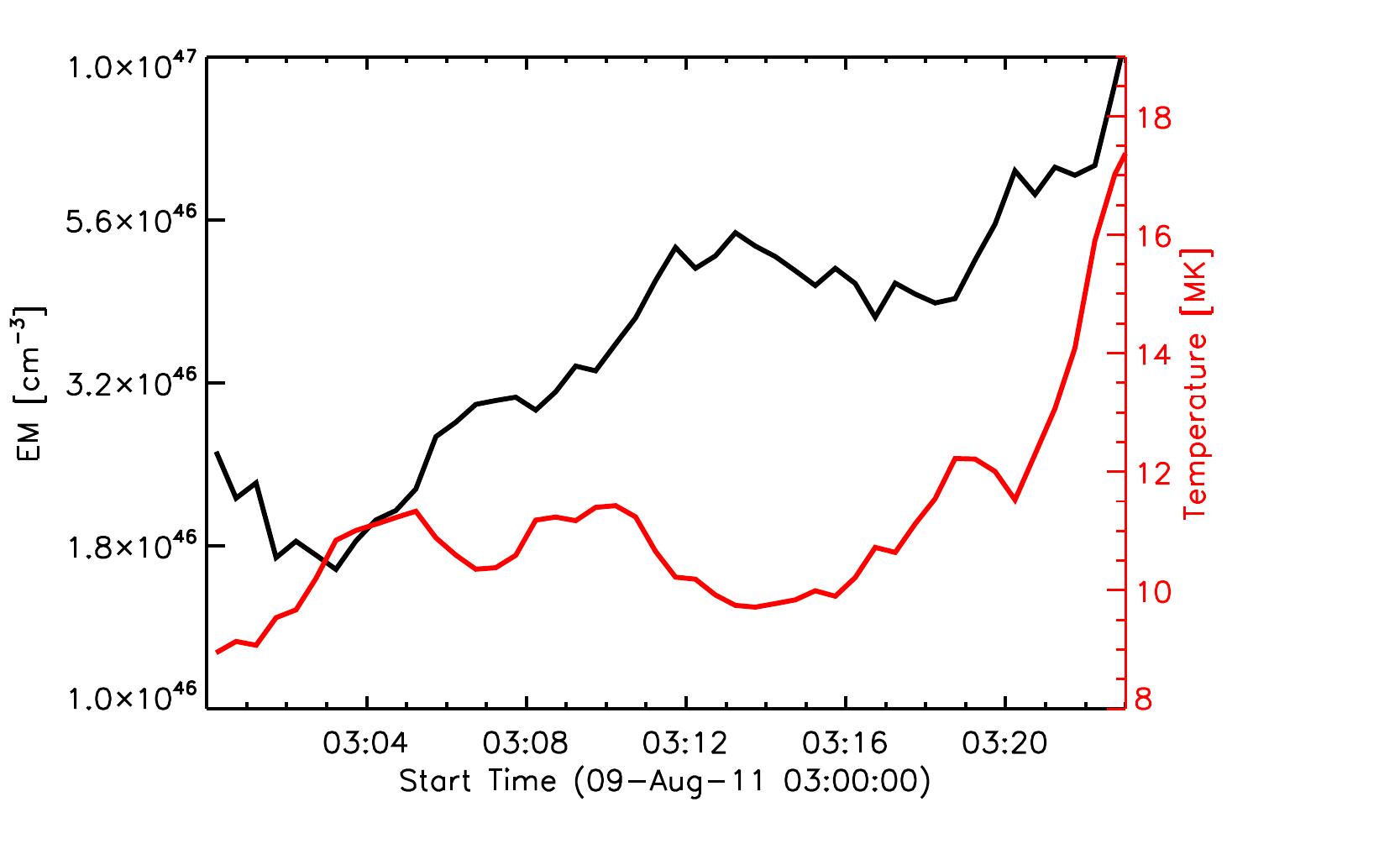} }
              \caption{Emission measure and temperature of preflare derived from RHESSI SXR data in the 6\,--\,12 keV energy range. {\it Black:} Emission measure; {\it red:} temperature.}
   \label{OSPEX}
   \end{figure}

  The SXR (6\,--\,12 keV) flux is shown in Figure \ref {light curve} with 4s time resolution. The light curve of the preflare phase shows three significant peaks. Their times are given in Table 1. The results of the OSPEX fitting with a 30s time integration are plotted in Figure \ref{OSPEX}. The three peaks in the SXR flux (Figure \ref {light curve}) are clearly noticeable in the time evolution of the temperature. The emission measure is time-variable as well, but with less pronounced peaks, which are delayed by 1\,--\,2 minutes. The delay of the emission measure relative to the temperature peak by several minutes is typical for microflare SXR emission \citep[{\it e.g.} ][]{2000SoPh..191..341K}. Figure \ref {light curve} suggests that the SXR preflare emission consists mostly of three peaks with some background that parallels the 17 GHz radio emission.

Table 1 presents the observed emission measure and temperature along with the inferred SXR luminosity for the three preflare peaks. The error of EM$_{\rm X}$ is dominated by the background subtraction. The SXR luminosity in the 0.1\,--\,10 keV range was calculated from EM$_{\rm X}$ and $T_{\rm X}$ using the APEC code \citep{2001ApJ...556L..91S}. Coronal elemental abundances, which are typical for flares \citep{2017ApJ...844...52D}, are assumed.

 \subsection{Radio Data and Analysis} 
  \label{Radio-text}
We have searched data from the Nobeyama Radioheliograph \citep[NoRH,][]{1994IEEEP..82..705N} at 17 GHz for non-thermal emission. In Figure \ref {light curve} the light curve is shown with 10 s time resolution. No significant peak is detected in the preflare phase. Thus we interpret the observed radio emission as purely thermal. As an upper limit for the non-thermal radio flux density $F_{\rm R}$ of the peaks we use 3 times the rms noise. A 20 s integration time was used for quantitative analysis. It yields an rms noise of $\sigma$=0.04 sfu (1 sfu = $10^{-19}$ erg s$^{-1}$ Hz$^{-1}$ cm$^{-2}$). Assuming isotropic emission,
\begin{equation}
  L_{\rm R} = 4 \pi D^2 F_{\rm R}\ \  ,
\end{equation}
where $D$ is the distance to the Sun. Equation 2 yields an upper limit radio luminosity density of $3.4 \times 10^7$ erg s$^{-1}$ Hz$^{-1}$. This value is also given for comparison in Table 1.

No enhanced radio emission in the 50\,--\,450 MHz range was reported during the preflare by the e-CALLISTO network (Ooty Observatory). Thus there is no evidence for non-thermal electrons at lower energies either.

\begin{table}

\begin{center}
\begin{tabular}{lcrccr}   
\hline\hline
Peak time&$EM_{\rm X}$& $T_{\rm X}$&$L_{\rm X}$& $L_{\rm R}$&\hskip-0.5cm log($L_{\rm X}/L_{\rm R}$) \\
hh:mm UT& [10$^{46}$cm$^{-3}$]&[MK]&[10$^{23}$ erg s$^{-1}$]& [10$^7$ erg s$^{-1}$ Hz$^{-1}$] &\\
\hline

03:05 & 2.0$\pm 1.2$& 11.3$\pm0.3$& 6.8& $<$ 3.4& $>$16.3\\
03:10 & 3.1$\pm 1.5$& 11.4$\pm0.3$& 10.0& $<$ 3.4&$>$16.5 \\
03:19 & 4.0$\pm 2.2$& 12.2$\pm0.3$& 12.6& $<$ 3.4&$>$16.7 \\
\hline
03:25 & 30.2$\pm 1.6$& 21.0$\pm0.1$& 57$\pm 13$&  590$\pm 60$& 15.0$\pm 0.1$\\
03:31 & 36.5$\pm 2.6$& 19.8$\pm0.1$& 70$\pm 13$&  390$\pm 30$& 15.2$\pm 0.1$\\
03:54 & 220$\pm 16$& 15.4$\pm0.1$& 513$\pm 36$&  1400$\pm 100$& 15.6$\pm 0.1$\\
\hline
\end{tabular}
\end{center}
\caption{ Preflare (upper part) and main flare phase (lower part) of SOL2011-08-09T03:52.} \label{table1}
\end{table}

\section{Discussion} 
      \label{S-discussion}

 \begin{figure}    
\centerline{\includegraphics[width=1.2\textwidth,clip=1]{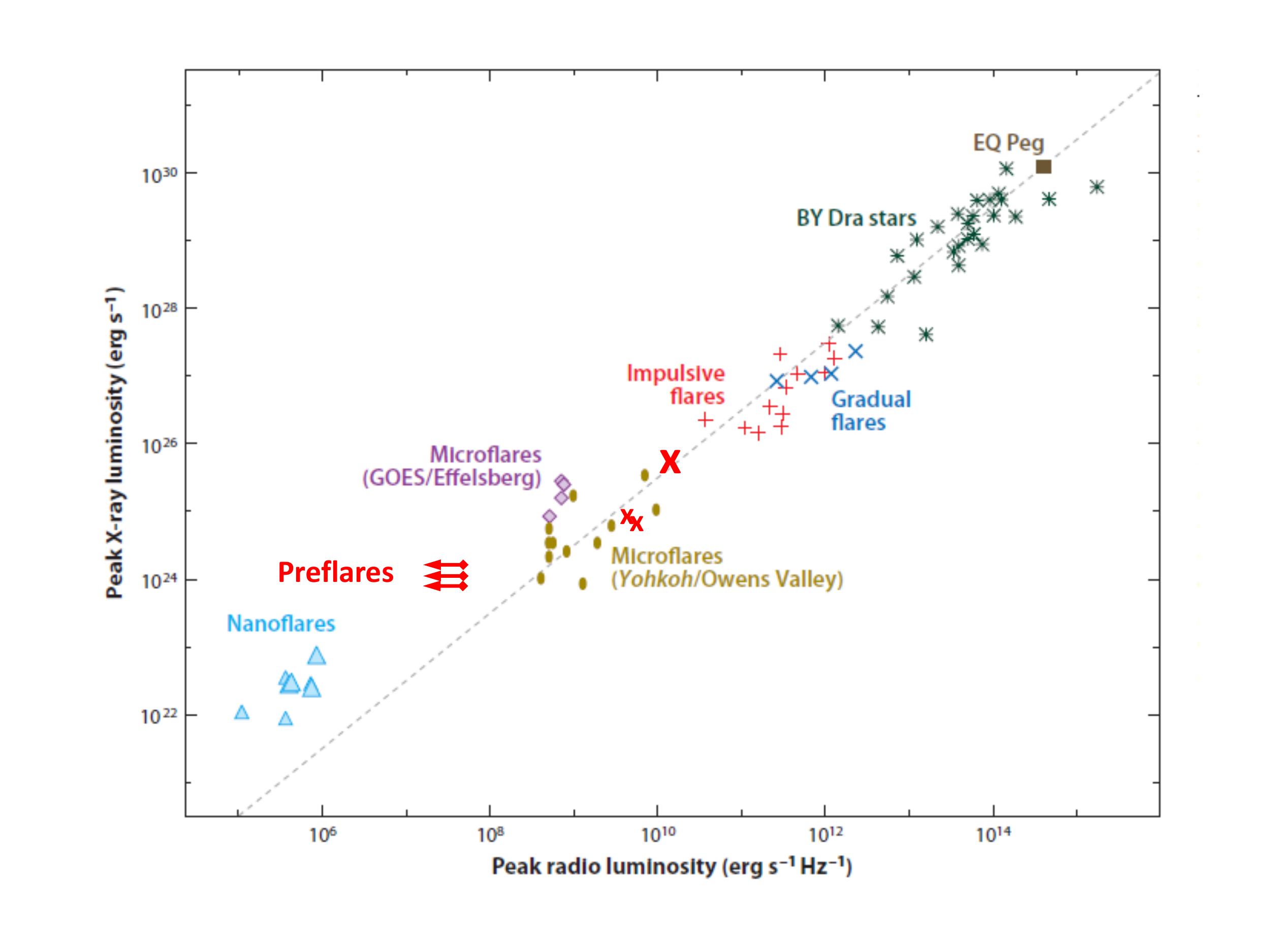} }
              \caption{Comparison of the peak soft X-ray and 3.6 cm radio luminosity of solar and stellar flares. Upper limits derived from these preflare observations are shown by red arrows. Two peaks in the early main flare phase are indicated with a {\it little red x}, the main peak with a {\it capital X}. Also shown are nanoflares in quiet regions \citep{2000SoPh..191..341K}, microflares in active regions \citep{1981Natur.291..210B, 1982A&A...107..178F, 1997ApJ...477..958G}, impulsive and gradual flares \citep{1994A&A...285..621B}, quiescent BY Dra, dMe, and dKe stars \citep{1993ApJ...415..236G}, and a long-duration flare on the dMe star EQ Peg \citep{1988A&A...195..159K}. The dashed line presents the relation given in Equation 1. (Graphics adapted from \citet{2010ARA&A..48..241B}, and based on \citet{2000SoPh..191..341K})}
   \label{comparison}
   \end{figure}

How do the luminosity ratios found here compare with typical ratios in solar flares?  Table 1 presents the ratios for the preflares (top), the two peaks recorded during the early phase of the main flare, and the main flare peak at 03:54 UT (bottom). The latter was not observed by RHESSI. So the measurements at 04:03 UT were used and interpolated back to the peak using GOES data. The preflare ratios are significantly higher. Figure \ref{comparison} shows a comparison of the presented values with values found for a large collection of flares, both on the Sun and on other stars. It can be seen that the ratio of SXR to radio emission in the preflare peaks presented here is at least an order of magnitude above that ratio in regular flares. The preflare peaks are radio-poor, resembling microflares in quiet regions (nanoflares).

The weakness of gyro-synchrotron emission may be the result of a weaker magnetic field in the source region or of an  underabundance of mildly relativistic electrons. Since the SXR preflare is observed at approximately the same location as the following regular flare, we prefer the second interpretation.

Feeble electron acceleration to high energies in the preflare phase complies with the spectral evolution of HXR emission in regular flares, which is initially steep and flattens during the maximum. The non-thermal electron distribution thus typically steepens again after the maximum and evolves according to the soft-hard-soft pattern. This property was noted already by \citet{1969ApJ...155L.117P} and in greater detail by \citet{2002ESASP.506..261H} and \citet{2004A&A...426.1093G}. Alternatively, but less frequent are flares that harden systematically during the event, thus showing a soft-hard-harder pattern \citep{1971ApJ...165..655F, 1986ApJ...305..920C, 1995ApJ...453..973K,2008ApJ...683.1180G}.

Low efficiency for acceleration of relativistic electrons in preflares is also consistent with the trend of small flares to have steeper (softer) non-thermal electron distributions at 35 keV than large flares \citep{2005A&A...439..737B}. On the average, the power-law spectral index increases from $-$7 to $-$3 for B6 to M5 flares. Similarly, microflares in quiet regions (nanoflares) have been reported to be radio-poor \citep{2000SoPh..191..341K}.

\section{Conclusions} 
      \label{S-conclusions}

We found three well developed peaks in the preflare SXR emission of the flare SOL2011-08-09T03:52 to have no corresponding radio emission. Upper limits on the radio flux are an order of magnitude below the level expected from regular flares relative to the observed SXR luminosity. The radio-poor events suggest the lack of efficient electron acceleration to high energies in the preflare phase. Gyro-synchrotron emission due to mildly relativistic electrons is particularly sensitive to this effect.

The observations are consistent with a scenario in which energy release in flares starts mainly with heating rather than acceleration in the preflare phase. Acceleration becomes dominant only in the main flare phase. Here heating is defined as an energization to a Maxwellian-like velocity distribution at higher temperature. We refer to acceleration if it results in a non-Maxwellian distribution such as a power-law.

\begin{acks}
We thank Paolo J.A. Sim{\~o}es for his help with the Nobeyama data and acknowledge the use of e-CALLISTO data (Christian Monstein, Institute for Astronomy, ETH Zurich and Institute for 4D Technologies, FHNW Windisch, Switzerland). Solar radio work at ETH was financially supported by the Swiss NSF (grants 20-113556 and 200020-121676), MB acknowledges support by SNF (grant 200021-140308) and MG by the FWF NFN projects S11601-N16 and S11604-N16.
\vskip2mm
\hskip-4mm
{\bf Disclosure of Potential Conflicts of Interest} The authors declare that they have no conflicts of interest.
\end{acks}

\bibliographystyle{spr-mp-sola}
\bibliography{preflareRX}

\begin{thebibliography}{36}
\ifx\bisbn     \undefined \def\bisbn  #1{ISBN #1}\fi
\ifx\binits    \undefined \def\binits#1{#1}\fi
\ifx\bauthor   \undefined \def\bauthor#1{#1}\fi
\ifx\batitle   \undefined \def\batitle#1{#1}\fi
\ifx\bjtitle   \undefined \def\bjtitle#1{\textit{#1}}\fi
\ifx\bvolume   \undefined \def\bvolume#1{\textbf{#1}}\fi
\ifx\byear     \undefined \def\byear#1{#1}\fi
\ifx\bissue    \undefined \def\bissue#1{#1}\fi
\ifx\bfpage    \undefined \def\bfpage#1{#1}\fi
\ifx\blpage    \undefined \def\blpage #1{#1}\fi
\ifx\burl      \undefined \def\burl#1{\textsf{#1}}\fi
\ifx\href      \undefined \def\href#1#2{\textsf{#2}}\fi
\ifx\betal     \undefined \def\betal{\textit{et al.}}\fi
\ifx\bctitle   \undefined \def\bctitle#1{#1}\fi
\ifx\beditor   \undefined \def\beditor#1{#1}\fi
\ifx\bbtitle   \undefined \def\bbtitle#1{\textit{#1}}\fi
\ifx\bedition  \undefined \def\bedition#1{#1}\fi
\ifx\bseriesno \undefined \def\bseriesno#1{\textbf{#1}}\fi
\ifx\blocation \undefined \def\blocation#1{#1}\fi
\ifx\bsertitle \undefined \def\bsertitle#1{\textit{#1}}\fi
\ifx\bsnm      \undefined \def\bsnm#1{#1}\fi
\ifx\bsuffix   \undefined \def\bsuffix#1{#1}\fi
\ifx\bparticle \undefined \def\bparticle#1{#1}\fi
\ifx\barticle  \undefined \def\barticle#1{}\fi
\ifx\binstitute  \undefined \def\binstitute#1{#1}\fi
\ifx\bpublisher  \undefined \def\bpublisher#1{#1}\fi
\ifx\doiurl    \undefined
  \def\doiurl#1{\href{http://dx.doi.org/#1}{\textsf{DOI}}}\fi
\ifx\arxivurl  \undefined
  \def\arxivurl#1{\href{http://arxiv.org/abs/#1}{\textsf{arXiv}}}\fi
\ifx\adsurl    \undefined
  \def\adsurl#1{\href{http://adsabs.harvard.edu/abs/#1}{\textsf{ADS}}}\fi
\ifx\botherref \undefined \def\botherref#1{}\fi
\ifx\url       \undefined \def\url#1{\textsf{#1}}\fi
\ifx\bchapter  \undefined \def\bchapter#1{}\fi
\ifx\bbook     \undefined \def\bbook#1{}\fi
\ifx\bcomment  \undefined \def\bcomment#1{#1}\fi
\ifx\oauthor   \undefined \def\oauthor#1{#1}\fi
\ifx\citeauthoryear \undefined\def \citeauthoryear#1{#1}\fi
\ifx\endbibitem\undefined \def\endbibitem{}\fi
\ifx\bconflocation  \undefined \def\bconflocation#1{#1} \fi

\bibitem[\protect\citeauthoryear{{Altyntsev}
  \textit{et~al.}}{2012}]{2012ApJ...758..138A}
\begin{barticle}
\bauthor{\bsnm{{Altyntsev}}, \binits{A.A.}},
\bauthor{\bsnm{{Fleishman}}, \binits{G.D.}},
\bauthor{\bsnm{{Lesovoi}}, \binits{S.V.}},
\bauthor{\bsnm{{Meshalkina}}, \binits{N.S.}}:
\byear{2012},
\batitle{{Thermal to Nonthermal Energy Partition at the Early Rise Phase of
  Solar Flares}}.
\bjtitle{\apj}
\bvolume{758},
\bfpage{138}.
\doiurl{10.1088/0004-637X/758/2/138}.
\adsurl{2012ApJ...758..138A}.
\end{barticle}
\endbibitem

\bibitem[\protect\citeauthoryear{{Aschwanden}
  \textit{et~al.}}{2017}]{2017ApJ...836...17A}
\begin{barticle}
\bauthor{\bsnm{{Aschwanden}}, \binits{M.J.}},
\bauthor{\bsnm{{Caspi}}, \binits{A.}},
\bauthor{\bsnm{{Cohen}}, \binits{C.M.S.}},
\bauthor{\bsnm{{Holman}}, \binits{G.}},
\bauthor{\bsnm{{Jing}}, \binits{J.}},
\bauthor{\bsnm{{Kretzschmar}}, \binits{M.}},
\bauthor{\bsnm{{Kontar}}, \binits{E.P.}},
\bauthor{\bsnm{{McTiernan}}, \binits{J.M.}},
\bauthor{\bsnm{{Mewaldt}}, \binits{R.A.}},
\bauthor{\bsnm{{O'Flannagain}}, \binits{A.}},
\bauthor{\bsnm{{Richardson}}, \binits{I.G.}},
\bauthor{\bsnm{{Ryan}}, \binits{D.}},
\bauthor{\bsnm{{Warren}}, \binits{H.P.}},
\bauthor{\bsnm{{Xu}}, \binits{Y.}}:
\byear{2017},
\batitle{{Global Energetics of Solar Flares. V. Energy Closure in Flares and
  Coronal Mass Ejections}}.
\bjtitle{\apj}
\bvolume{836},
\bfpage{17}.
\doiurl{10.3847/1538-4357/836/1/17}.
\adsurl{2017ApJ...836...17A}.
\end{barticle}
\endbibitem

\bibitem[\protect\citeauthoryear{{Bastian}, {Benz}, and
  {Gary}}{1998}]{1998ARA&A..36..131B}
\begin{barticle}
\bauthor{\bsnm{{Bastian}}, \binits{T.S.}},
\bauthor{\bsnm{{Benz}}, \binits{A.O.}},
\bauthor{\bsnm{{Gary}}, \binits{D.E.}}:
\byear{1998},
\batitle{{Radio Emission from Solar Flares}}.
\bjtitle{\araa}
\bvolume{36},
\bfpage{131}.
\doiurl{10.1146/annurev.astro.36.1.131}.
\adsurl{1998ARA\%26A..36..131B}.
\end{barticle}
\endbibitem

\bibitem[\protect\citeauthoryear{{Battaglia}, {Fletcher}, and
  {Benz}}{2009}]{2009A&A...498..891B}
\begin{barticle}
\bauthor{\bsnm{{Battaglia}}, \binits{M.}},
\bauthor{\bsnm{{Fletcher}}, \binits{L.}},
\bauthor{\bsnm{{Benz}}, \binits{A.O.}}:
\byear{2009},
\batitle{{Observations of conduction driven evaporation in the early rise phase
  of solar flares}}.
\bjtitle{\aap}
\bvolume{498},
\bfpage{891}.
\doiurl{10.1051/0004-6361/200811196}.
\adsurl{2009A\%26A...498..891B}.
\end{barticle}
\endbibitem

\bibitem[\protect\citeauthoryear{{Battaglia}, {Fletcher}, and
  {Sim{\~o}es}}{2014}]{2014ApJ...789...47B}
\begin{barticle}
\bauthor{\bsnm{{Battaglia}}, \binits{M.}},
\bauthor{\bsnm{{Fletcher}}, \binits{L.}},
\bauthor{\bsnm{{Sim{\~o}es}}, \binits{P.J.A.}}:
\byear{2014},
\batitle{{Where is the Chromospheric Response to Conductive Energy Input from a
  Hot Pre-flare Coronal Loop?}}
\bjtitle{\apj}
\bvolume{789},
\bfpage{47}.
\doiurl{10.1088/0004-637X/789/1/47}.
\adsurl{2014ApJ...789...47B}.
\end{barticle}
\endbibitem

\bibitem[\protect\citeauthoryear{{Battaglia}, {Grigis}, and
  {Benz}}{2005}]{2005A&A...439..737B}
\begin{barticle}
\bauthor{\bsnm{{Battaglia}}, \binits{M.}},
\bauthor{\bsnm{{Grigis}}, \binits{P.C.}},
\bauthor{\bsnm{{Benz}}, \binits{A.O.}}:
\byear{2005},
\batitle{{Size dependence of solar X-ray flare properties}}.
\bjtitle{\aap}
\bvolume{439},
\bfpage{737}.
\doiurl{10.1051/0004-6361:20053027}.
\adsurl{2005A\%26A...439..737B}.
\end{barticle}
\endbibitem

\bibitem[\protect\citeauthoryear{{Benz}}{2017}]{2017LRSP...14....2B}
\begin{barticle}
\bauthor{\bsnm{{Benz}}, \binits{A.O.}}:
\byear{2017},
\batitle{{Flare Observations}}.
\bjtitle{Living Reviews in Solar Physics}
\bvolume{14},
\bfpage{2}.
\doiurl{10.1007/s41116-016-0004-3}.
\adsurl{2017LRSP...14....2B}.
\end{barticle}
\endbibitem

\bibitem[\protect\citeauthoryear{{Benz} and
  {G\"udel}}{1994}]{1994A&A...285..621B}
\begin{barticle}
\bauthor{\bsnm{{Benz}}, \binits{A.O.}},
\bauthor{\bsnm{{G\"udel}}, \binits{M.}}:
\byear{1994},
\batitle{{X-ray/microwave ratio of flares and coronae}}.
\bjtitle{\aap}
\bvolume{285},
\bfpage{621}.
\adsurl{1994A\%26A...285..621B}.
\end{barticle}
\endbibitem

\bibitem[\protect\citeauthoryear{{Benz} and
  {G{\"u}del}}{2010}]{2010ARA&A..48..241B}
\begin{barticle}
\bauthor{\bsnm{{Benz}}, \binits{A.O.}},
\bauthor{\bsnm{{G{\"u}del}}, \binits{M.}}:
\byear{2010},
\batitle{{Physical Processes in Magnetically Driven Flares on the Sun, Stars,
  and Young Stellar Objects}}.
\bjtitle{\araa}
\bvolume{48},
\bfpage{241}.
\doiurl{10.1146/annurev-astro-082708-101757}.
\adsurl{2010ARA\%26A..48..241B}.
\end{barticle}
\endbibitem

\bibitem[\protect\citeauthoryear{{Benz}
  \textit{et~al.}}{1981}]{1981Natur.291..210B}
\begin{barticle}
\bauthor{\bsnm{{Benz}}, \binits{A.O.}},
\bauthor{\bsnm{{Perrenoud}}, \binits{M.R.}},
\bauthor{\bsnm{{F\"urst}}, \binits{E.}},
\bauthor{\bsnm{{Hirth}}, \binits{W.}}:
\byear{1981},
\batitle{{Solar radio blips and X-ray kernels}}.
\bjtitle{\nat}
\bvolume{291},
\bfpage{210}.
\doiurl{10.1038/291210a0}.
\adsurl{1981Natur.291..210B}.
\end{barticle}
\endbibitem

\bibitem[\protect\citeauthoryear{{Benz}
  \textit{et~al.}}{1983}]{1983SoPh...83..267B}
\begin{barticle}
\bauthor{\bsnm{{Benz}}, \binits{A.O.}},
\bauthor{\bsnm{{Barrow}}, \binits{C.H.}},
\bauthor{\bsnm{{Dennis}}, \binits{B.R.}},
\bauthor{\bsnm{{Pick}}, \binits{M.}},
\bauthor{\bsnm{{Raoult}}, \binits{A.}},
\bauthor{\bsnm{{Simnett}}, \binits{G.}}:
\byear{1983},
\batitle{{X-ray and radio emissions in the early stages of solar flares}}.
\bjtitle{\solphys}
\bvolume{83},
\bfpage{267}.
\doiurl{10.1007/BF00148280}.
\adsurl{1983SoPh...83..267B}.
\end{barticle}
\endbibitem

\bibitem[\protect\citeauthoryear{{Cliver}
  \textit{et~al.}}{1986}]{1986ApJ...305..920C}
\begin{barticle}
\bauthor{\bsnm{{Cliver}}, \binits{E.W.}},
\bauthor{\bsnm{{Dennis}}, \binits{B.R.}},
\bauthor{\bsnm{{Kiplinger}}, \binits{A.L.}},
\bauthor{\bsnm{{Kane}}, \binits{S.R.}},
\bauthor{\bsnm{{Neidig}}, \binits{D.F.}},
\bauthor{\bsnm{{Sheeley}}, \binits{N.R.} \bsuffix{Jr.}},
\bauthor{\bsnm{{Koomen}}, \binits{M.J.}}:
\byear{1986},
\batitle{{Solar gradual hard X-ray bursts and associated phenomena}}.
\bjtitle{\apj}
\bvolume{305},
\bfpage{920}.
\doiurl{10.1086/164306}.
\adsurl{1986ApJ...305..920C}.
\end{barticle}
\endbibitem

\bibitem[\protect\citeauthoryear{{Dennis} and
  {Zarro}}{1993}]{1993SoPh..146..177D}
\begin{barticle}
\bauthor{\bsnm{{Dennis}}, \binits{B.R.}},
\bauthor{\bsnm{{Zarro}}, \binits{D.M.}}:
\byear{1993},
\batitle{{The Neupert effect - What can it tell us about the impulsive and
  gradual phases of solar flares?}}
\bjtitle{\solphys}
\bvolume{146},
\bfpage{177}.
\doiurl{10.1007/BF00662178}.
\adsurl{1993SoPh..146..177D}.
\end{barticle}
\endbibitem

\bibitem[\protect\citeauthoryear{{Emslie}
  \textit{et~al.}}{2012}]{2012ApJ...759...71E}
\begin{barticle}
\bauthor{\bsnm{{Emslie}}, \binits{A.G.}},
\bauthor{\bsnm{{Dennis}}, \binits{B.R.}},
\bauthor{\bsnm{{Shih}}, \binits{A.Y.}},
\bauthor{\bsnm{{Chamberlin}}, \binits{P.C.}},
\bauthor{\bsnm{{Mewaldt}}, \binits{R.A.}},
\bauthor{\bsnm{{Moore}}, \binits{C.S.}},
\bauthor{\bsnm{{Share}}, \binits{G.H.}},
\bauthor{\bsnm{{Vourlidas}}, \binits{A.}},
\bauthor{\bsnm{{Welsch}}, \binits{B.T.}}:
\byear{2012},
\batitle{{Global Energetics of Thirty-eight Large Solar Eruptive Events}}.
\bjtitle{\apj}
\bvolume{759},
\bfpage{71}.
\doiurl{10.1088/0004-637X/759/1/71}.
\adsurl{2012ApJ...759...71E}.
\end{barticle}
\endbibitem

\bibitem[\protect\citeauthoryear{{Frost} and
  {Dennis}}{1971}]{1971ApJ...165..655F}
\begin{barticle}
\bauthor{\bsnm{{Frost}}, \binits{K.J.}},
\bauthor{\bsnm{{Dennis}}, \binits{B.R.}}:
\byear{1971},
\batitle{{Evidence from Hard X-Rays for Two-Stage Particle Acceleration in a
  Solar Flare}}.
\bjtitle{\apj}
\bvolume{165},
\bfpage{655}.
\doiurl{10.1086/150932}.
\adsurl{1971ApJ...165..655F}.
\end{barticle}
\endbibitem

\bibitem[\protect\citeauthoryear{{F\"urst}, {Benz}, and
  {Hirth}}{1982}]{1982A&A...107..178F}
\begin{barticle}
\bauthor{\bsnm{{F\"urst}}, \binits{E.}},
\bauthor{\bsnm{{Benz}}, \binits{A.O.}},
\bauthor{\bsnm{{Hirth}}, \binits{W.}}:
\byear{1982},
\batitle{{About the relation between radio and soft X-ray emission in case of
  very weak solar activity}}.
\bjtitle{\aap}
\bvolume{107},
\bfpage{178}.
\adsurl{1982A\%26A...107..178F}.
\end{barticle}
\endbibitem

\bibitem[\protect\citeauthoryear{{Gary}, {Hartl}, and
  {Shimizu}}{1997}]{1997ApJ...477..958G}
\begin{barticle}
\bauthor{\bsnm{{Gary}}, \binits{D.E.}},
\bauthor{\bsnm{{Hartl}}, \binits{M.D.}},
\bauthor{\bsnm{{Shimizu}}, \binits{T.}}:
\byear{1997},
\batitle{{Nonthermal Radio Emission from Solar Soft X-Ray Transient
  Brightenings}}.
\bjtitle{\apj}
\bvolume{477},
\bfpage{958}.
\doiurl{10.1086/303748}.
\adsurl{1997ApJ...477..958G}.
\end{barticle}
\endbibitem

\bibitem[\protect\citeauthoryear{{Grigis} and
  {Benz}}{2004}]{2004A&A...426.1093G}
\begin{barticle}
\bauthor{\bsnm{{Grigis}}, \binits{P.C.}},
\bauthor{\bsnm{{Benz}}, \binits{A.O.}}:
\byear{2004},
\batitle{{The spectral evolution of impulsive solar X-ray flares}}.
\bjtitle{\aap}
\bvolume{426},
\bfpage{1093}.
\doiurl{10.1051/0004-6361:20041367}.
\adsurl{2004A\%26A...426.1093G}.
\end{barticle}
\endbibitem

\bibitem[\protect\citeauthoryear{{Grigis} and
  {Benz}}{2008}]{2008ApJ...683.1180G}
\begin{barticle}
\bauthor{\bsnm{{Grigis}}, \binits{P.C.}},
\bauthor{\bsnm{{Benz}}, \binits{A.O.}}:
\byear{2008},
\batitle{{Spectral Hardening in Large Solar Flares}}.
\bjtitle{\apj}
\bvolume{683},
\bfpage{1180}.
\doiurl{10.1086/589826}.
\adsurl{2008ApJ...683.1180G}.
\end{barticle}
\endbibitem

\bibitem[\protect\citeauthoryear{{G\"udel} and
  {Benz}}{1993}]{1993ApJ...405L..63G}
\begin{barticle}
\bauthor{\bsnm{{G\"udel}}, \binits{M.}},
\bauthor{\bsnm{{Benz}}, \binits{A.O.}}:
\byear{1993},
\batitle{{X-ray/microwave relation of different types of active stars}}.
\bjtitle{\apjl}
\bvolume{405},
\bfpage{L63}.
\doiurl{10.1086/186766}.
\adsurl{1993ApJ...405L..63G}.
\end{barticle}
\endbibitem

\bibitem[\protect\citeauthoryear{{G\"udel}
  \textit{et~al.}}{1993}]{1993ApJ...415..236G}
\begin{barticle}
\bauthor{\bsnm{{G\"udel}}, \binits{M.}},
\bauthor{\bsnm{{Schmitt}}, \binits{J.H.M.M.}},
\bauthor{\bsnm{{Bookbinder}}, \binits{J.A.}},
\bauthor{\bsnm{{Fleming}}, \binits{T.A.}}:
\byear{1993},
\batitle{{A tight correlation between radio and X-ray luminosities of M
  dwarfs}}.
\bjtitle{\apj}
\bvolume{415},
\bfpage{236}.
\doiurl{10.1086/173158}.
\adsurl{1993ApJ...415..236G}.
\end{barticle}
\endbibitem

\bibitem[\protect\citeauthoryear{{Hudson} and
  {F{\'a}rn{\'{\i}}k}}{2002}]{2002ESASP.506..261H}
\begin{bchapter}
\bauthor{\bsnm{{Hudson}}, \binits{H.S.}},
\bauthor{\bsnm{{F{\'a}rn{\'{\i}}k}}, \binits{F.}}:
\byear{2002},
\bctitle{{Spectral variations of flare hard X-rays}}.
In: \beditor{\bsnm{{Wilson}}, \binits{A.}} (ed.)
\bbtitle{Solar Variability: From Core to Outer Frontiers},
\bsertitle{ESA Special Publication}
\bseriesno{506},
\bfpage{261}.
\adsurl{2002ESASP.506..261H}.
\end{bchapter}
\endbibitem

\bibitem[\protect\citeauthoryear{{Kiplinger}}{1995}]{1995ApJ...453..973K}
\begin{barticle}
\bauthor{\bsnm{{Kiplinger}}, \binits{A.L.}}:
\byear{1995},
\batitle{{Comparative Studies of Hard X-Ray Spectral Evolution in Solar Flares
  with High-Energy Proton Events Observed at Earth}}.
\bjtitle{\apj}
\bvolume{453},
\bfpage{973}.
\doiurl{10.1086/176457}.
\adsurl{1995ApJ...453..973K}.
\end{barticle}
\endbibitem

\bibitem[\protect\citeauthoryear{{Kosugi}, {Dennis}, and
  {Kai}}{1988}]{1988ApJ...324.1118K}
\begin{barticle}
\bauthor{\bsnm{{Kosugi}}, \binits{T.}},
\bauthor{\bsnm{{Dennis}}, \binits{B.R.}},
\bauthor{\bsnm{{Kai}}, \binits{K.}}:
\byear{1988},
\batitle{{Energetic electrons in impulsive and extended solar flares as deduced
  from flux correlations between hard X-rays and microwaves}}.
\bjtitle{\apj}
\bvolume{324},
\bfpage{1118}.
\doiurl{10.1086/165967}.
\adsurl{1988ApJ...324.1118K}.
\end{barticle}
\endbibitem

\bibitem[\protect\citeauthoryear{{Krucker} and
  {Benz}}{2000}]{2000SoPh..191..341K}
\begin{barticle}
\bauthor{\bsnm{{Krucker}}, \binits{S.}},
\bauthor{\bsnm{{Benz}}, \binits{A.O.}}:
\byear{2000},
\batitle{{Are Heating Events in the Quiet Solar Corona Small Flares?
  Multiwavelength Observations of Individual Events}}.
\bjtitle{\solphys}
\bvolume{191},
\bfpage{341}.
\doiurl{10.1023/A:1005255608792}.
\adsurl{2000SoPh..191..341K}.
\end{barticle}
\endbibitem

\bibitem[\protect\citeauthoryear{{Kundu}
  \textit{et~al.}}{1988}]{1988A&A...195..159K}
\begin{barticle}
\bauthor{\bsnm{{Kundu}}, \binits{M.R.}},
\bauthor{\bsnm{{White}}, \binits{S.M.}},
\bauthor{\bsnm{{Jackson}}, \binits{P.D.}},
\bauthor{\bsnm{{Pallavicini}}, \binits{R.}}:
\byear{1988},
\batitle{{Co-ordinated VLA and EXOSAT observations of the flare stars UV Ceti,
  EQ Pegasi, YZ Canis Minoris, and AD Leonis}}.
\bjtitle{\aap}
\bvolume{195},
\bfpage{159}.
\adsurl{1988A\%26A...195..159K}.
\end{barticle}
\endbibitem

\bibitem[\protect\citeauthoryear{{Lin}}{2011}]{2011SSRv..159..421L}
\begin{barticle}
\bauthor{\bsnm{{Lin}}, \binits{R.P.}}:
\byear{2011},
\batitle{{Energy Release and Particle Acceleration in Flares: Summary and
  Future Prospects}}.
\bjtitle{\ssr}
\bvolume{159},
\bfpage{421}.
\doiurl{10.1007/s11214-011-9801-0}.
\adsurl{2011SSRv..159..421L}.
\end{barticle}
\endbibitem

\bibitem[\protect\citeauthoryear{{Lin}
  \textit{et~al.}}{2002}]{2002SoPh..210....3L}
\begin{barticle}
\bauthor{\bsnm{{Lin}}, \binits{R.P.}},
\bauthor{\bsnm{{Dennis}}, \binits{B.R.}},
\bauthor{\bsnm{{Hurford}}, \binits{G.J.}},
\bauthor{\bsnm{{Smith}}, \binits{D.M.}},
\bauthor{\bsnm{{Zehnder}}, \binits{A.}},
\bauthor{\bsnm{{Harvey}}, \binits{P.R.}},
\bauthor{\bsnm{{Curtis}}, \binits{D.W.}},
\bauthor{\bsnm{{Pankow}}, \binits{D.}},
\bauthor{\bsnm{{Turin}}, \binits{P.}},
\bauthor{\bsnm{{Bester}}, \binits{M.}},
\bauthor{\bsnm{{Csillaghy}}, \binits{A.}},
\bauthor{\bsnm{{Lewis}}, \binits{M.}},
\bauthor{\bsnm{{Madden}}, \binits{N.}},
\bauthor{\bsnm{{van Beek}}, \binits{H.F.}},
\bauthor{\bsnm{{Appleby}}, \binits{M.}},
\bauthor{\bsnm{{Raudorf}}, \binits{T.}},
\bauthor{\bsnm{{McTiernan}}, \binits{J.}},
\bauthor{\bsnm{{Ramaty}}, \binits{R.}},
\bauthor{\bsnm{{Schmahl}}, \binits{E.}},
\bauthor{\bsnm{{Schwartz}}, \binits{R.}},
\bauthor{\bsnm{{Krucker}}, \binits{S.}},
\bauthor{\bsnm{{Abiad}}, \binits{R.}},
\bauthor{\bsnm{{Quinn}}, \binits{T.}},
\bauthor{\bsnm{{Berg}}, \binits{P.}},
\bauthor{\bsnm{{Hashii}}, \binits{M.}},
\bauthor{\bsnm{{Sterling}}, \binits{R.}},
\bauthor{\bsnm{{Jackson}}, \binits{R.}},
\bauthor{\bsnm{{Pratt}}, \binits{R.}},
\bauthor{\bsnm{{Campbell}}, \binits{R.D.}},
\bauthor{\bsnm{{Malone}}, \binits{D.}},
\bauthor{\bsnm{{Landis}}, \binits{D.}},
\bauthor{\bsnm{{Barrington-Leigh}}, \binits{C.P.}},
\bauthor{\bsnm{{Slassi-Sennou}}, \binits{S.}},
\bauthor{\bsnm{{Cork}}, \binits{C.}},
\bauthor{\bsnm{{Clark}}, \binits{D.}},
\bauthor{\bsnm{{Amato}}, \binits{D.}},
\bauthor{\bsnm{{Orwig}}, \binits{L.}},
\bauthor{\bsnm{{Boyle}}, \binits{R.}},
\bauthor{\bsnm{{Banks}}, \binits{I.S.}},
\bauthor{\bsnm{{Shirey}}, \binits{K.}},
\bauthor{\bsnm{{Tolbert}}, \binits{A.K.}},
\bauthor{\bsnm{{Zarro}}, \binits{D.}},
\bauthor{\bsnm{{Snow}}, \binits{F.}},
\bauthor{\bsnm{{Thomsen}}, \binits{K.}},
\bauthor{\bsnm{{Henneck}}, \binits{R.}},
\bauthor{\bsnm{{McHedlishvili}}, \binits{A.}},
\bauthor{\bsnm{{Ming}}, \binits{P.}},
\bauthor{\bsnm{{Fivian}}, \binits{M.}},
\bauthor{\bsnm{{Jordan}}, \binits{J.}},
\bauthor{\bsnm{{Wanner}}, \binits{R.}},
\bauthor{\bsnm{{Crubb}}, \binits{J.}},
\bauthor{\bsnm{{Preble}}, \binits{J.}},
\bauthor{\bsnm{{Matranga}}, \binits{M.}},
\bauthor{\bsnm{{Benz}}, \binits{A.}},
\bauthor{\bsnm{{Hudson}}, \binits{H.}},
\bauthor{\bsnm{{Canfield}}, \binits{R.C.}},
\bauthor{\bsnm{{Holman}}, \binits{G.D.}},
\bauthor{\bsnm{{Crannell}}, \binits{C.}},
\bauthor{\bsnm{{Kosugi}}, \binits{T.}},
\bauthor{\bsnm{{Emslie}}, \binits{A.G.}},
\bauthor{\bsnm{{Vilmer}}, \binits{N.}},
\bauthor{\bsnm{{Brown}}, \binits{J.C.}},
\bauthor{\bsnm{{Johns-Krull}}, \binits{C.}},
\bauthor{\bsnm{{Aschwanden}}, \binits{M.}},
\bauthor{\bsnm{{Metcalf}}, \binits{T.}},
\bauthor{\bsnm{{Conway}}, \binits{A.}}:
\byear{2002},
\batitle{{The Reuven Ramaty High-Energy Solar Spectroscopic Imager (RHESSI)}}.
\bjtitle{\solphys}
\bvolume{210},
\bfpage{3}.
\doiurl{10.1023/A:1022428818870}.
\adsurl{2002SoPh..210....3L}.
\end{barticle}
\endbibitem

\bibitem[\protect\citeauthoryear{{Nakajima}
  \textit{et~al.}}{1994}]{1994IEEEP..82..705N}
\begin{barticle}
\bauthor{\bsnm{{Nakajima}}, \binits{H.}},
\bauthor{\bsnm{{Nishio}}, \binits{M.}},
\bauthor{\bsnm{{Enome}}, \binits{S.}},
\bauthor{\bsnm{{Shibasaki}}, \binits{K.}},
\bauthor{\bsnm{{Takano}}, \binits{T.}},
\bauthor{\bsnm{{Hanaoka}}, \binits{Y.}},
\bauthor{\bsnm{{Torii}}, \binits{C.}},
\bauthor{\bsnm{{Sekiguchi}}, \binits{H.}},
\bauthor{\bsnm{{Bushimata}}, \binits{T.}},
\bauthor{\bsnm{{Kawashima}}, \binits{S.}},
\bauthor{\bsnm{{Shinohara}}, \binits{N.}},
\bauthor{\bsnm{{Irimajiri}}, \binits{Y.}},
\bauthor{\bsnm{{Koshiishi}}, \binits{H.}},
\bauthor{\bsnm{{Kosugi}}, \binits{T.}},
\bauthor{\bsnm{{Shiomi}}, \binits{Y.}},
\bauthor{\bsnm{{Sawa}}, \binits{M.}},
\bauthor{\bsnm{{Kai}}, \binits{K.}}:
\byear{1994},
\batitle{{The Nobeyama radioheliograph.}}
\bjtitle{IEEE Proceedings}
\bvolume{82},
\bfpage{705}.
\adsurl{1994IEEEP..82..705N}.
\end{barticle}
\endbibitem

\bibitem[\protect\citeauthoryear{{Neupert}}{1968}]{1968ApJ...153L..59N}
\begin{barticle}
\bauthor{\bsnm{{Neupert}}, \binits{W.M.}}:
\byear{1968},
\batitle{{Comparison of Solar X-Ray Line Emission with Microwave Emission
  during Flares}}.
\bjtitle{\apjl}
\bvolume{153},
\bfpage{L59}.
\doiurl{10.1086/180220}.
\adsurl{1968ApJ...153L..59N}.
\end{barticle}
\endbibitem

\bibitem[\protect\citeauthoryear{{Parks} and
  {Winckler}}{1969}]{1969ApJ...155L.117P}
\begin{barticle}
\bauthor{\bsnm{{Parks}}, \binits{G.K.}},
\bauthor{\bsnm{{Winckler}}, \binits{J.R.}}:
\byear{1969},
\batitle{{Sixteen-Second Periodic Pulsations Observed in the Correlated
  Microwave and Energetic X-Ray Emission from a Solar Flare}}.
\bjtitle{\apjl}
\bvolume{155},
\bfpage{L117}.
\doiurl{10.1086/180315}.
\adsurl{1969ApJ...155L.117P}.
\end{barticle}
\endbibitem

\bibitem[\protect\citeauthoryear{{Priest}}{2014}]{2014masu.book.....P}
\begin{bbook}
\bauthor{\bsnm{{Priest}}, \binits{E.}}:
\byear{2014},
\bbtitle{{Magnetohydrodynamics of the Sun}}
\bseriesno{82},
\bpublisher{Cambridge University Press}, \blocation{Cambridge UK},
\bfpage{705}.
\adsurl{2014masu.book.....P}.
\end{bbook}
\endbibitem

\bibitem[\protect\citeauthoryear{{Shibata} and
  {Magara}}{2011}]{2011LRSP....8....6S}
\begin{barticle}
\bauthor{\bsnm{{Shibata}}, \binits{K.}},
\bauthor{\bsnm{{Magara}}, \binits{T.}}:
\byear{2011},
\batitle{{Solar Flares: Magnetohydrodynamic Processes}}.
\bjtitle{Living Reviews in Solar Physics}
\bvolume{8},
\bfpage{6}.
\doiurl{10.12942/lrsp-2011-6}.
\adsurl{2011LRSP....8....6S}.
\end{barticle}
\endbibitem

\bibitem[\protect\citeauthoryear{{Siarkowski}, {Falewicz}, and
  {Rudawy}}{2009}]{2009ApJ...705L.143S}
\begin{barticle}
\bauthor{\bsnm{{Siarkowski}}, \binits{M.}},
\bauthor{\bsnm{{Falewicz}}, \binits{R.}},
\bauthor{\bsnm{{Rudawy}}, \binits{P.}}:
\byear{2009},
\batitle{{Plasma Heating in the Very Early Phase of Solar Flares}}.
\bjtitle{\apjl}
\bvolume{705},
\bfpage{L143}.
\doiurl{10.1088/0004-637X/705/2/L143}.
\adsurl{2009ApJ...705L.143S}.
\end{barticle}
\endbibitem

\bibitem[\protect\citeauthoryear{{Smith}
  \textit{et~al.}}{2001}]{2001ApJ...556L..91S}
\begin{barticle}
\bauthor{\bsnm{{Smith}}, \binits{R.K.}},
\bauthor{\bsnm{{Brickhouse}}, \binits{N.S.}},
\bauthor{\bsnm{{Liedahl}}, \binits{D.A.}},
\bauthor{\bsnm{{Raymond}}, \binits{J.C.}}:
\byear{2001},
\batitle{{Collisional Plasma Models with APEC/APED: Emission-Line Diagnostics
  of Hydrogen-like and Helium-like Ions}}.
\bjtitle{\apjl}
\bvolume{556},
\bfpage{L91}.
\doiurl{10.1086/322992}.
\adsurl{2001ApJ...556L..91S}.
\end{barticle}
\endbibitem

\bibitem[\protect\citeauthoryear{{Zharkova}
  \textit{et~al.}}{2011}]{2011SSRv..159..357Z}
\begin{barticle}
\bauthor{\bsnm{{Zharkova}}, \binits{V.V.}},
\bauthor{\bsnm{{Arzner}}, \binits{K.}},
\bauthor{\bsnm{{Benz}}, \binits{A.O.}},
\bauthor{\bsnm{{Browning}}, \binits{P.}},
\bauthor{\bsnm{{Dauphin}}, \binits{C.}},
\bauthor{\bsnm{{Emslie}}, \binits{A.G.}},
\bauthor{\bsnm{{Fletcher}}, \binits{L.}},
\bauthor{\bsnm{{Kontar}}, \binits{E.P.}},
\bauthor{\bsnm{{Mann}}, \binits{G.}},
\bauthor{\bsnm{{Onofri}}, \binits{M.}},
\bauthor{\bsnm{{Petrosian}}, \binits{V.}},
\bauthor{\bsnm{{Turkmani}}, \binits{R.}},
\bauthor{\bsnm{{Vilmer}}, \binits{N.}},
\bauthor{\bsnm{{Vlahos}}, \binits{L.}}:
\byear{2011},
\batitle{{Recent Advances in Understanding Particle Acceleration Processes in
  Solar Flares}}.
\bjtitle{\ssr}
\bvolume{159},
\bfpage{357}.
\doiurl{10.1007/s11214-011-9803-y}.
\adsurl{2011SSRv..159..357Z}.
\end{barticle}
\endbibitem

\end{thebibliography}

\end{article}
\end{document}